\documentclass[11pt,aps,showpacs,showkeys]{revtex4}
\usepackage{amsmath,amsthm,amssymb,epsfig,alltt}

\begin{document}

\def\tr{{\rm tr}}
\def\a{\alpha}
\def\b{\beta}
\def\d{{\delta}}
\def\l{\lambda}
\def\e{\epsilon}
\def\p{\partial}
\def\m{\mu}
\def\n{\nu}
\def\t{\tau}
\def\th{\theta}
\def\s{\sigma}
\def\g{\gamma}
\def\G{\Gamma}
\def\o{\omega}
\def\r{\rho}
\def\D{\Delta}
\def\half{\frac{1}{2}}
\def\hatt{{\hat t}}
\def\hatx{{\hat x}}
\def\hatp{{\hat p}}
\def\hatX{{\hat X}}
\def\hatY{{\hat Y}}
\def\hatP{{\hat P}}
\def\haty{{\hat y}}
\def\whatX{{\widehat{X}}}
\def\whata{{\widehat{\alpha}}}
\def\whatb{{\widehat{\beta}}}
\def\whatV{{\widehat{V}}}
\def\hatth{{\hat \theta}}
\def\hatta{{\hat \tau}}
\def\hatrh{{\hat \rho}}
\def\hatva{{\hat \varphi}}
\def\barx{{\bar x}}
\def\bary{{\bar y}}
\def\barz{{\bar z}}
\def\baro{{\bar \omega}}
\def\barpsi{{\bar \psi}}
\def\sp{\sigma^\prime}
\def\nn{\nonumber}
\def\cb{{\cal B}}
\def\2pap{2\pi\alpha^\prime}
\def\wideA{\widehat{A}}
\def\wideF{\widehat{F}}
\def\beq{\begin{eqnarray}}
 \def\eeq{\end{eqnarray}}
 \def\4pap{4\pi\a^\prime}
 \def\xp{{x^\prime}}
 \def\sp{{\s^\prime}}
 \def\ap{{\a^\prime}}
 \def\tp{{\t^\prime}}
 \def\zp{{z^\prime}}
 \def\xpp{x^{\prime\prime}}
 \def\xppp{x^{\prime\prime\prime}}
 \def\barxp{{\bar x}^\prime}
 \def\barxpp{{\bar x}^{\prime\prime}}
 \def\barxppp{{\bar x}^{\prime\prime\prime}}
 \def\barchi{{\bar \chi}}
 \def\baro{{\bar \omega}}
 \def\bpsi{{\bar \psi}}
 \def\barg{{\bar g}}
 \def\barz{{\bar z}}
 \def\bareta{{\bar \eta}}
 \def\ta{{\tilde \a}}
 \def\tb{{\tilde \b}}
 \def\tc{{\tilde c}}
 \def\tz{{\tilde z}}
 \def\tJ{{\tilde J}}
 \def\tpsi{\tilde{\psi}}
 \def\tal{{\tilde \alpha}}
 \def\tbe{{\tilde \beta}}
 \def\tga{{\tilde \gamma}}
 \def\tchi{{\tilde{\chi}}}
 \def\barth{{\bar \theta}}
 \def\bareta{{\bar \eta}}
 \def\barom{{\bar \omega}}
 \def\bole{{\boldsymbol \epsilon}}
 \def\bolth{{\boldsymbol \theta}}
 \def\bomega{{\boldsymbol \omega}}
 \def\bolmu{{\boldsymbol \mu}}
 \def\bola{{\boldsymbol \alpha}}
 \def\bolb{{\boldsymbol \beta}}
 \def\bolX{{\boldsymbol X}}
 \def\mathN{{\boldsymbol n}}
 \def\bba{{\boldsymbol a}}
 \def\bbA{{\boldsymbol A}}
 \def\mathP{{\mathbb P}}
 \def\mathN{{\boldsymbol N}}
 \def\mathN{{\mathbb N}}
 \def\bbP{{\boldsymbol P}}

\setcounter{page}{1}
\title[]{NSR Open Super-string in the proper-time gauge I: Free Field Theory
}

\author{Taejin Lee}
\affiliation{
Department of Physics, Kangwon National University, Chuncheon 24341
Korea}

\email{taejin@kangwon.ac.kr}

\begin{abstract}

We study the Neveu-Schwarz-Ramond (NSR) open super-string theory in the proper-time gauge. The string field action is obtained by evaluating the Polyakov string path integral. In this study, we focus on the open-string free-field action, which corresponds to the string path integral on a strip. Depending on the periodicity of the fermion fields, the open super-string has two sectors: The Neveu-Schwarz (NS) and Ramond (R) sectors. We can impose the gauge conditions to fix the (super-) reparametrization invariance on the two-dimensional metric and its super-partner on the string world sheet to secure the covariance, in contrast to the light-cone gauge condition. 
Accordingly, the proper-time emerges in the NS sector and 
both proper-time and its super-partner appear in the R-sector. Integration leads to free-string field actions in both sectors. 

\end{abstract}




\maketitle

\newpage
\tableofcontents
\newpage

\section{Introduction}

Quantum field theory describes the dynamics of nature in terms of the quantum 
fields of point particles. Since its introduction by Dirac \cite{Dirac}, it has been the language in which one has attempted to understand fundamental forces. Quantum field theory flourished as quantum electrodynamics and quantum chromodynamics, which describes the strong interaction, and electroweak theory in the 1970s, which combines electromagnetism and weak interaction in a unified scheme. 

Currently, we are confronting a new challenge of constructing quantum field theories of strings and replacing point particles with strings as fundamental objects. For the super-string, two approaches to construct quantum field theory are available: the light-cone field theory of the NSR super-string \cite{mandelstam1974a,Green1983a,Green1983b,Mandelstam1986,Grossperi1987,Sin1989} and Witten's super-string field theory based on the extended BRST symmetry \cite{Witten1986nucl}. Each approach helps us to understand the super-string dynamics and mechanism of the super-string field theory to a great extent; however, both approaches have limitations. 
The light-cone super-string theory is not manifestly covariant and is plagued with various divergences \cite{Greensite1987a,Greensite1987b,Greensite1988,Green1988,Ishibashi2017}. Witten's super-string field theory also suffers from divergence owing to the mid-point contact interaction \cite{Wendt1989} .

In this study, we propose a new covariant approach based on the Polyakov string path integral \cite{Polyakov1981}. Because the Polyakov string path integral is well-defined and finite, we expect that the divergence issues associated with the other two approaches may be resolved using this approach. 
When we evaluate the Polyakov string path integral on a strip of string worksheet, we can obtain a covariant field theoretical propagator if we impose the gauge condition to fix the reparametrization invariance. This approach has been applied to the bosonic string theory \cite{Lee88,TLeeJKPS2017,Lee2017d,TLee2017cov,TLeeEPJ2018,Lai2017S,TLee2019PLB}. An important advantage of this approach is that it is easy to include higher interaction terms, and we can easily evaluate three -and four-string interactions to confirm local gauge invariance.
The interacting NSR superstring in the proper-time gauge may be free of the notorious picture changing problem of conformal field theory formulation, if properly developed. As I put as the sub-title of the paper, "Free Field Theory", this paper will serve as a preliminary to a work \cite{TLeeprep} in this direction.

As a first step toward constructing a covariant interacting super-string theory, in the present study, we will focus on the free-field action of NSR super-strings. 
To obtain the free-field action, we evaluate the Polyakov string path integral on a strip with two spatial boundaries. By applying canonical quantization, we determine that the Hamiltonian only comprises constraints. To secure the covariance, we impose the gauge condition on the world-sheet metric and its super-partner. Depending on the periodicity of the fermion fields, the NSR super-string has two sectors: periodic in the NS sector and antiperiodic in the Ramond sector. We demonstrate that the NS sector has a bosonic modular parameter that 
becomes the proper time, and the R sector has both bosonic modular and fermionic supermodular parameters. After integrating both the modular and super-modular parameters, we obtain a Dirac propagator in the Ramond sector. In the NS sector, we obtain a Klein-Gordon type propagator, integrating the modular parameter (proper time).

\section{Canonical Quantization of NSR Super-String}

The reparametrization-invariant and local super-symmetric actions for the NSR super-string are given by: 
\beq
I &=& \int d^2 \xi L, \nn\\
L &=& \sqrt{-h} \Biggl\{ - \half h^{\a\b} \frac{\p X^\m}{\p \xi^\a} \frac{\p X_\m}{\p \xi^\b}- \frac{i}{2} \bar \psi^\m \g^a \frac{\p \psi_\m}{\p \xi^\a} + \frac{1}{2}F^\m F_\m + \frac{1}{2} \bar \chi_\a \g^\b \frac{\p X^\m}{\p \xi^\b} \g^\a \psi_\m \nn\\
&&+\frac{1}{16} \bar \psi^\m \psi_\m \bar \chi_\a \g^\b \g^\a \chi_\b
\Biggr\} 
\eeq 
An auxiliary field $F^\m$ is introduced to render the local super-symmetric algebra closed off-shell. The action is invariant under the reparametrization given by 
\beq
\d \xi&=& \zeta^a, ~~ \d X^\m = \zeta^\a \p_\a X^\m, ~~ \d \psi^\m = \zeta^\a \p_\a \psi^\m , \nn\\
\d e_\a{}^a &=& \zeta^\b \p_\b e_\a{}^a + e_\b{}^a \p_\a \zeta^\b ,~~
\d \xi_\a = \zeta^\b \p_\b \chi_a + \chi_\b \p_\a \zeta^\b 
\eeq 
and local super-symmetric transformation given by 
\beq
\d X^\m =&=& i \e \psi^\m, ~~ \d \psi^\m = \g^\a \left(\p_\a X^\m - \frac{1}{2} \bar \chi_a \psi^\m \right) \e, \nn\\
\d e_\a{}^a &=& \e \g^\a \chi_a, ~~ \d \chi_a = 2i D_\a \e 
\eeq 
where the covariant derivative $D_\a$ is defined by 
\beq
D_\a e = \p_\a \e - \frac{1}{2} \o_\a \g^5 \e,
\eeq 
with the connection 
\beq
\o_a = - \frac{1}{\e} e_\a{}^a \e^{\b\g} \p_\b e_\g{}^b \eta_{ab} + \frac{1}{2} \bar \chi_\a \g^5 \g^\b \chi_\b . 
\eeq 
In addition to the local super-symmetric transformation and reparametrization, the action is laso-invariant under conformal transformation. 
\beq
\d X^\m =0, ~~ \d \psi^\m = - \frac{1}{2} \e \psi^\m, ~~ \d h_{\a\b} = 2 \e h_{\a\b}, ~~ \d \chi_a = \frac{1}{2} \e \chi_\a 
\eeq 
and the super-conformal transformation 
\beq
\d X^\m = 0, ~~ \d \psi^\m = 0, ~~ \d h_{\a\b} =0, ~~ \d \chi^\a = \g^\a \e .
\eeq 
At the critical dimensions, we set the conformal factor to $e^\phi =1$ and 
fix the super-conformal invariance, setting 
\beq
\g^\a \chi_\a = 0.
\eeq 

To construct the Hamiltonian describing the dynamics of the string , an extended one-dimensional object, we express the metric $h_{\a\b}$ in terms of the lapse and shift functions
\beq
h^{\a\b} = \frac{1}{N_1} \left(\begin{array}{cc} -1 & N_2 \\ 
N_2 & (N_1)^2 -(N_2)^2\end{array} \right) . 
\eeq 
Accordingly, zweibein $e_\a{}^a$ is in terms of the lapse, and the shift functions are expressed as 
\beq
\left(e_\a{}^a\right) = \frac{1}{\sqrt{N_1}} \left(\begin{array}{cc} 1 & - N_2 \\ 0 & N_1 \end{array} \right). 
\eeq 
Two orthogonal vectors on the world-sheet are given as 
\beq
{\boldsymbol e}_{\hat 0} &=& \frac{1}{\sqrt{N_1}} \left({\boldsymbol \p}_\t - N_2 {\boldsymbol \p}_\s \right) \nn\\
{\boldsymbol e}_{\hat 1} &=& \sqrt{N_1} {\boldsymbol \p}_\s
\eeq 

The Hamiltonian can be obtained by taking the Legendre transformation of the Lagrangian and defining the canonical conjugates $(P^\m, \Pi^\a, \zeta)$ to $(X_\m, N_\a, \l = -\sqrt{2} N^{\frac{5}{4}}_1 \bar \chi_1)$: 
\beq
P^\m = \frac{\p L}{\p \dot X_\m}, ~~ \Pi^\a = \frac{\p L}{\p \dot N_a},~~
\zeta = \frac{\p L}{\p \dot \l}.
\eeq 
From the defining equations of the momenta, we obtain 
\beq
P^\m = \frac{1}{N_1} (\dot X^\m - N_2 X^{\prime \m}) + \bar \chi^0 \psi^\m,
\eeq 
and the first class primary constraints
\beq
\Pi_\a = 0, ~~\zeta = 0.
\eeq 
With some algebra, we determine
\beq
\bar H &=& P^\m \dot X_\m -L \nn\\
&=& \frac{N_1}{2} (P^2 + X^{\prime 2}) +N_2 P^\m X^\prime_\m + \frac{\l}{\sqrt{2}}(\g^5 P^\m + X^{\prime \m}) N^{-\frac{1}{4}}_1 \psi_\m +
\frac{1}{2} \bar \psi^\m \g^\a \frac{\p}{\p \xi^\a} \psi_\m . 
\eeq 
Here, we adopt a simple two- dimensional $\g$ matrix algebra $\g^\a \g^\b \g_\a =0$, the Fierz rearrangement
\beq
\bar \psi_a \psi_2 \psi_3 = - \frac{1}{2} \sum_i \bar \psi_1 \G_i \psi_3 \G^i \psi_2, ~~~ \G_i = 1, \g_a, \g_5 .
\eeq 
We need to scale $\psi \rightarrow N^{-\frac{1}{4}}_1 \psi^\m$ to express the fermion in canonical form: 
\beq
H &=& \frac{N_1}{2} \left(P^2+ X^{\prime 2} + i \psi^\m \g^5 \psi^{\prime}_\m\right) + N_2 \left( P^\m X^\prime_\m + \frac{i}{2} \psi^\m \psi^\prime_\m \right) \nn\\
&&
+ \frac{\l}{\sqrt{2}} \left(\g^5 P^\m + X^{\prime \m} \right) \psi_\m - v_\a \Pi^\a - u \zeta
\eeq 
where Lagrangian multipliers $v_\a$ and $u$ are introduced to enforce the primary constraints. The primary constraints $\Pi_\a =0$ and $\zeta=0$ lead to the secondary constraints given as follows:
\beq
P^2 + X^{\prime 2}+ i \psi^\m \g^5 \psi^\prime_\m &=&0, \nn\\
P^\m X^\prime_\m + \frac{i}{2} \psi^\m \psi^\prime_\m &=& 0, \\
\frac{1}{\sqrt{2}} \left(\g^5 P^\m + X^{\prime \m} \right) \psi_\m &=& 0. \nn
\eeq

\section{NSR Open super-string}

We may extend the domain of $\s$, initially defined as $[0,\pi]$ for the open super-string, to $[-\pi, \pi]$, such that the field variables for the open string and those for the closed super-string are defined in the same domain. Furthermore, for the open super-string, we impose the following folding conditions on the dynamical variables:
\beq
X^\m(\s) &=& X^\m(-\s), ~~P^\m(\s) = p^\m(-\s), ~~ N_1(\s) = N_1(-\s), ~~ N_2(\s) = - N_2(-\s), \nn\\
\Pi_1(\s) &=& \Pi_1(-\s), ~~\Pi_2(\s) = - \Pi_2(-\s),~~ \psi^{\m+}(\s) =\psi^{\m-}(-\s),~~\l^{\m+}(\s) =-\l^{\m-}(-\s)
\eeq 
The conditions are also read in terms of notmal modes as follows
\beq
X^\m_n &=& X^\m_{-n}, ~~ P^\m_n = P^\m_{-n}, ~~ N^1_n =N^1_{-n},~~ N^2_n = - N^2_{-n} 
\eeq 
and 
\beq
\begin{cases} \psi^{\m+}_n = \psi^{\m-}_n ,~~ \l^+_{n} = - \l^-_{-n} ~~~~~& \text{for the Ramond sector,} \\
\psi^{\m+}_{n+\half} = \psi^{\m-}_{n+\half},~~ \l^+_{n+\half} = - \l^-_{-n-\half} ~~~~~& \text{for the Neuveu-Schwarz sector.} 
\end{cases}
\eeq 
With these conditions, we can write action $S$ in the Ramond sector as follows: 
\beq
S &=& \int^{\t_f}_{\t_i} d\t \sum_n \Biggl\{ P^\m_n \dot X_{\m n} + \frac{i}{2} \psi^\m_{-n} \dot \psi_{\m n} + \bar \Pi_n \dot {\bar N}_n + \bar \zeta_n \dot {\bar\l}_n - \bar N_n L^R_n \nn\\
&& - \bar \l_n F_n + v_n \bar \Pi_n + u_n \bar \zeta_n \Biggr\},
\eeq 
where in the Ramond sector
\beq
\bar N_n &=& \frac{1}{2} \left(N_{1n} + N_{2n}\right), ~~ \bar N_0 = N_10, ~~ \bar \Pi_n = \frac{1}{2} (\Pi_{1n} + \Pi_{2n} ), ~~ \bar \Pi_0 = \Pi_{10}, n\not=0, \nn\\
 \bar \l_n &=& \l^+_n + \l^-_n, ~~ \bar \zeta_n = \zeta^+_n + \zeta^-_n ,
\eeq
and in the Neveu-Schwarz sector
\beq
S &=& \int^{\t_f}_{\t_i} d\t \sum_n \Biggl\{ P^\m_n \dot X_{\m n} + \frac{i}{2} \psi^\m_{-n-\half} \dot \psi_{\m n+\half} + \bar \Pi_n \dot {\bar N}_n + \bar \zeta_{n+\half} \dot {\bar\l}1_{n+\half} - \bar N_n L^R_n \nn\\
&& - \bar \l_{n+\half} G_{n+\half} + v_n \bar \Pi_n + u_{n+\half} \bar \zeta_{n+\half} \Biggr\},
\eeq 
with
\beq
\bar N_n &=& \frac{1}{2} \left(N_{1n} + N_{2n}\right), ~~ \bar N_0 = N_10, ~~ \bar \Pi_n = \frac{1}{2} (\Pi_{1n} + \Pi_{2n} ), ~~ \bar \Pi_0 = \Pi_{10}, n\not=0, \nn\\
\bar \l_{n+\half} &=& \l^+_{n+\half} + \l^-_{n+\half}, ~~ \bar \zeta_{n+\half} = \zeta^+_{n+\half} + \zeta^-_{n+\half} ,
\eeq

\section{Covariant Proper-Time Gauge condition for the NSR Open Super-String}

\subsubsection{Neveu-Schwarz Sector} 

In the Neveu-Schwarz sector, the secondary constraints $L^R_n$, $G_{n+\half}$ form the super-Virasoro algebra with the central charge 
\beq \label{nsconst}
\left[L^{NS}_n, L^{NS}_m \right] &=& (n-m) L^{NS}_{n+m} + \frac{d}{8} n (n^2-1) \d(n+m), \nn\\
\left[ L^{LS}_n, G_{m+\half}\right ] &=& \left(\frac{n}{2} -n - \half \right) G_{n+m+\half}, \\
\left[G_{n+\half}, G_{m-\half} \right] &=& 2 L^{NS}_{n+m} + \frac{d}{2} n(n+1) \d(n+m). \nn
\eeq 
The canonical generator that generates the reparametrization and local supersymmetric transformation is constructed to be in the NS sector. 
\beq
\Omega_{NS}(\e, \o) &=& \dot \e_n \bar Pi_n + \e_n \Biggl\{ L^{NS}_n -i(2n-j) \bar N_{j-n} \bar \Pi_j +i \left(\frac{3}{2}n -j - \half \right) \bar \l_{j-n+\half} \bar\zeta_{j+\half}\Biggr\} \nn\\
&&+ \dot \o_{n+\half} \bar \zeta_{n+\half} - \o_{n+\half} \Biggl\{G_{n+\half} + 2i \bar \l_{j-n-\half} \bar \Pi_j - \frac{i}{2} (3n-j+1) \bar N_{j-n} \bar \zeta_{j+\half} \Biggl\}
\eeq 
To secure the Lorentz covariance, it is desirable to impose gauge fixing conditions on the Lagrangian multipliers $\bar N_n$ and $\bar \l_{n+\half}$. The Lagrangian multipliers transform the gauge transformations by 
\beq
\d \bar N_n &=& \dot \e_n + i(n-2m) \e_m \bar N_{n-m} -2 i \o_{m+\half} \bar \l_{n-m-\half}, \nn\\
\d \bar \l_{n+\half} &=& - \dot \o_{n+\half} + i \left(n-\frac{3}{2}n + \half \right) \e_m \bar \l_{n-m+\half} - \frac{i}{2}(n-3m-1)\o_{m+\half} \bar N_{n-m}.
\eeq 
We may choose the covariant gauge condition for the NSR open super-string in the Neveu-Schwarz sector by 
\beq \label{gaugefixing}
\bar N_{n(\not=0)} &=&0, ~~ \dot {\bar N}_0 = 0, \nn\\
\bar \l_{n+\half} &=& 0 .
\eeq 

We can confirm that this proper-time gauge fixes the gauge degrees of freedom associated with the reparametrization and local super-symmetry completely and consistently. Near the gauge-fixing hypersurface defined by Eq. (\ref{gaugefixing}), the infinitesimal gauge transformation that restores the gauge-fixing condition along the gauge orbits is determined by 
\beq
\dot \e_n - in \e_n \bar n + \bar N_n &=& 0, ~~~n \not=0, \nn\\
\ddot \e_0 + \bar N_0 &=&0, \\
\dot \o_{n+\half} - \left(n+\half\right) \o_{n+\half} \bar n - \bar \l_{n+\half} &=&0 \nn 
\eeq 
where $\bar n$ denotes constant $\bar N_0$ in the covariant gauge. This linear differential equation for $\e_n$ and $\o_{n+\half}$ has a unique solution:
\beq
\e_n(\t) &=& e^{in \bar n \t} \left\{ - \int^\t_{\t_i}e^{-in \bar n \t^\prime} \bar N_n dt^\prime + c_n \right\}, ~~ n \not=0, \nn\\
\e_0(\t) &=& \left\{ \frac{\t-\t_i)}{T} \int^{\t_f}_\t + \frac{\t-\t_f)}{T} \int^{\t}_{\t_i} \right\} \bar N_0 d\t^\prime, ~~T = \t_f - \t_i \\
\o_{n+\half} &=& e^{(n+\half)\bar n \t }\left\{\int^\t_{\t_i} e^{-i(n+\half)\bar n \t^\prime} \bar \l_{n+\half} d \t^\prime + d_{n+\half}\right\} \nn
\eeq 
where 
\beq
c_n &=& \left(1- e^{-2in \bar n T} \right)^{-1} \left\{\int^{\t_f}_{\t_i}
e^{-in \bar n \t^\prime} \bar N_n d \t^\prime + e^{-2in \bar n \t_f }
\int^{\t_f}_{\t_i} e^{in \bar n \t^\prime} \bar N_{-n} d \t^\prime 
\right\} \nn\\
d_{n+\half} &=&\left(1- e^{-2i(n+ \half) \bar n T} \right)^{-1} \Biggl\{ e^{-2i(n+ \half)\bar n \t_f}\int^{\t_f}_{\t_i},
e^{i(n+ \half) \bar n \t^\prime} \bar \l_{-n-\half} d \t^\prime \nn\\
&& -
\int^{\t_f}_{\t_i} e^{-i(n+ \half) \bar n \t^\prime} \bar \l_{n+ \half} d \t^\prime \Biggr\} .
\eeq 

\subsubsection{Ramond Sector} 

In the Ramons sector, the constraint operators form a super-Virasoro algebra given by 
\beq \label{constramond}
\left[L^R_n, L^R_m\right] &=& (n-m)L^R_{n+m} + \frac{d}{8} n^3 \d(n+m), \nn\\
\left[L^R_n, F_m \right] &=& \left(\frac{n}{2}-m \right) F_{n+m}, \\
\left[F_n, F_m\right] &=& 2L^R_{n+m} + \frac{d}{2} n^2 \d (n+m). \nn
\eeq
The canonical generator of symmetric transformation is obtained in the Ramond sector as
\beq
\Omega_R(\e,\o) &=& \dot \e_n \bar \Pi_n + \e_n \left\{L^R_n -i(2n-j) \bar N_{j-n} \bar \Pi_j + i \left(\frac{3n}{2}-i\right) \bar \l_{j-n}\bar \zeta_j\right\} \nn\\ &&+ \dot \o_n \bar \zeta_n - \o_n \left\{F_n + 2i \bar \l_{j-n} \bar \Pi_j + i \left(\frac{3n}{2}- \frac{j}{2}\right) \bar N_{j-n}\bar \zeta_j \right\}. 
\eeq
We observe that the Lagrangian multipliers, $\bar N_n$ and $\bar \l_n$, transform under the representation and local supersymmetric transformation by 
\beq
\d \bar N_n &=& \dot \e_n + i (n-2m) \e_m \bar N_{n-m} - 2i\o_m \bar \l_{n-m} \nn\\
\d \bar \l_n &=& - \dot \o_n + i \left(n-\frac{3m}{2} \right) \e_m \bar \l_{n-m} - \frac{i}{2}(n-3m) \o_m \bar N_{n-m}.
\eeq 
We choose the covariant gauge condition in the Ramond sector by 
\beq \label{cov2}
\bar N_n = 0, ~~ \dot {\bar N}_0 = 0 ,~~ \bar \l_n =0, ~~ \dot {\bar \l}_0 =0, ~~ n \not=0.
\eeq 
The covariant gauge condition in Eq. (\ref{cov2}) properly fixes gauge symmetry, we examine the equation for infinitesimal gauge parameters that restore the gauge-fixing condition along the gauge orbits 
near gauge fixing hyper-surface by 
\beq \label{gaugeeq2}
\dot \e_n - in \e_n \bar n - 2i \o_n \n + \bar N_n &=& 0, \nn \\
\dot \o_n + \frac{n}{2} i \e_n \n - in \o_n \bar n - \bar \l_n &=& 0, \nn\\
\ddot \e_0 - 2i \dot \o_0 \n + \dot {\bar N}_0 &=& 0, \nn\\
\ddot \o_0 - \dot {\bar \l}_0 &=& 0, \nn
\eeq 
where $n \not =0$,$\bar n$, and $\n$ denote $\bar N_0$ and $\bar \l_0$, respectively. The existence of a unique solution to Eqs. (\ref{gaugeeq2}) ensures that the chosen covariant gauge condition is complete and consistent. With some algebra, we explicitly determine a unique solution:
\beq
\e_n(\t) &=& e^{in \bar n \t} \left\{- \int^\t_{\t_i} e^{-in \bar n \t^\prime}
(\bar N_n - 2i\o_n \n ) d\t^\prime + c_n \right\}, \nn\\
\e_0(\t) &=& \left\{\frac{\t-\t_i)}{T} \int^{\t_f}_{\t} + \frac{(\t-\t_f)}{T}\int^\t_{\t_f} \right\} (\bar N_0-2i\o_0 \n ) d\t^\prime, \nn\\
\o_n(\t) &=& e^{in \bar n \t} \left\{\int^\t_{\t_i} e^{-in\bar n \t^\prime}( \bar \l_n - \frac{n}{2} i \e_n \n) d \t^\prime + d_n \right\} \nn\\
\o_0(\t) &=& \left\{\frac{(\t_i -\t)}{T} \int^{\t_f}_\t + \frac{(\t_f-\t)}{T} \int^\t_{\t_i} \right\} \bar \l_0 d \t^\prime, ~~n\not=0.
\eeq 
where 
\beq
c_n &=& \left(1-e^{-2in \bar n T}\right)^{-1}\Biggl\{\int^{\t_f}_{\t_i} e^{-in\bar n \t^\prime} (\bar N_n -2i \o_n \n) d \t^\prime \nn\\
&&+ e^{-2in \bar n \t_f} \int^{\t_f}_{\t_i} e^{in\bar n \t^\prime} (\bar N_n -2i \o_{-n} \n ) d \t^\prime\Biggr\} \nn\\
d_n &=&\left(1-e^{-2in \bar n T}\right)^{-1}\Biggl\{
e^{-2in \bar n \t_f} \int^{\t_f}_{\t_i} e^{in\bar n \t^\prime} (\bar \l_{-n} + \frac{n}{2} i \e_{-n} \n) d \t^\prime \nn\\
&&
-\int^{\t_f}_{\t_i} e^{-in\bar n \t^\prime} (\bar \l_n -\frac{n}{2}i \o_n \n) d \t^\prime \Biggr\} 
\eeq

In contrast to the Neveu-Schwarz sector, we cannot gauge away the fermionic zero mode of the Lagrangian multipliers $\bar \l_0$ completely. This leads to 
the main difference between the two sectors is that in the Ramond sector, modular and super-modular parameters exist, while only modular parameters exist in the Neveu-Schwarz sector.

\section{The Off-Shell Propagator for the NSR Open Super-String}

In this section, we apply BRST quantization to the NSR open superstring. The path integral adopted to represent the off-shell propagator is evaluated explicitly. 

\subsection{Neveu-Schwarz Sector} 

First, we construct the BRST generator $Q_{NS}$ with the given structure constants in Eq. (\ref{nsconst}), introducing the fermionic ghost variables $\eta_n$, $\bar \eta_n$, $\xi_n$, and the bosonic ghost variables $\b_{n+\half}, \bar \b_{n+\half}, \g_{n+\half}$
\beq
Q_{NS} &=& \eta_n L^{NS}_n + \bar \eta_n \bar \Pi_n + \b_{n+\half} G_{n+\half} +\bar\b_{n+\half} \bar \zeta_{n+\half} - \frac{1}{2}(n-m) \eta_n \eta_m \xi_{n+m} \nn, \\
&& + \left(\frac{n}{2}-m -\half\right) \eta_n \b_{m+\half} \g_{n+m+\half} - \b_{n+\half} \b_{m-\half} \xi_{n+m},
\eeq 
(the BRST ghost variables $\eta_n$ and $\xi_n$ may be identified in terms of the usual $b$–$c$ ghost variables as 
$\eta_n = c_{-n},$ and $\xi_n = b_n$.)

Second, the BRST invariant effective action is constructed as 
\beq
S &=& \int^{\t_f}_{\t_i} d\t \Bigl\{ P^\m \dot X_{\m n} + \frac{i}{2} \psi^\m_{-n-\half} \dot \psi_{\m (n+\half)} + \bar \Pi_n \dot {\bar N}_n + \bar \zeta_{n + \half} \dot {\bar \l}_{n+\half} + i \xi_n \dot \eta_n + i \bar \xi_n \dot {\bar \eta}_n \nn, \\
&& +i \g_{n+\half} \dot \b_{n+\half} + i \bar \g_{n+\half} \dot {\bar \b}_{n+\half} -i [Q_{NS}, \Delta] \Bigr\},
\eeq 
Here we choose $\Delta$ by 
\beq
\Delta = \xi_n \bar N_n + \bar \eta_n \chi_n - \g_{n+\half} \bar \l_{n+\half} - \bar \g_{n+\half} f_{n+\half}, 
\eeq 
with 
\beq
\chi_n = \frac{1}{\a} \bar N_n, (n\not=0), \chi_0 = 0, ~~ f_{n+\half} = \frac{1}{\a} \bar \l_{n+\half} 
\eeq 
to produce covariant gauge conditions at limit $\a \rightarrow 0$. Scaling of dynamic variables 
\beq
(\bar \Pi_n, \bar \eta_n) &\rightarrow& \a (\bar \Pi_n, \bar \eta_n), ~~ n\not=0, \nn\\
(\bar \zeta_{n+\half}, \bar \b_{n+\half} ) &\rightarrow& \a (\bar \zeta_{n+\half}, \bar \b_{n+\half} ), 
\eeq 
we find , $\a \rightarrow 0$ in the limit, 
\beq
S &=& \int^{\t_f}_{\t_i} d\t \Biggl\{\sum_n \left(P^\m_n \dot X_{\m n} + \frac{i}{2} \psi^\m_{-n-\half} \dot \psi_{\m n+ \half} + i \xi_n \dot \eta_n +i \g_{n+\half} \dot \b_{n+\half}\right) \nn\\
&& + \sum_n \left(\bar N_n L^{NS}_n- \dot{\bar \l}_{n+\half} G_{n+\half} \right) -\sum_{n,m} \Bigl((n-m) \eta_n \bar N_m \xi_{n+m} \nn, \\
&&+ \left(\frac{n}{2}-m -\half\right) \bar N_n \b_{m+\half} \g_{n+m+\half}
+ \left(\frac{n}{2}-m-\half\right) \eta_n \bar \l_{m+\half} \g_{n+m+\half} \nn\\
&&+ 2\b_{n+\half} \bar \l_{m-\half} \xi_{n+m} \Bigr)-\sum_n \left(\bar \zeta_{n+\half} \bar \l_{n+\half}+ i\bar\b_{n+\half} \bar \g_{n+\half} \right), \nn\\.
&& - \sum_n{}^\prime (\bar \Pi_n N_n -i \bar \eta_n \xi_n ) + \bar \Pi_0 \dot{\bar N}_0 + i \bar \xi \dot{\bar \eta}_0 -i\xi_0 \bar \eta_0 \Biggr\}
\eeq 
We integrate the conjugates of Lagrangian multipliers $\bar \Pi_n$ and $\bar \zeta_{n+\half}$, and obtain the action parameterized by modular parameter $\bar n$. Ghost variables $(\bar \eta_n, \bar \xi_n)$, $n\not=0$, and $(\bar \b_{n+\half}, \bar \g_{n+\half})$ can be trivially integrated. By further integrating over the ghost zero modes $(\eta_0, \xi_0)$ and $(\bar \eta_0, \bar \xi_0)$, we obtain the factor 
\beq
Det[\p^2_\t] = T.
\eeq 
Finally, we obtain the resultant action as 
\beq
S &=& \int^{\t_f}_{\t_i} d\t \Biggl\{\sum_n \left(P^\m_n \dot X_{\m n} + \frac{i}{2} \psi^\m_{-n-\half} \dot \psi_{\m n+ \half}+i \g_{n+\half} \dot \b_{n+\half}\right)
\nn\\
&& +i \sum_n{}^\prime \xi_n \dot \eta_n - \bar n \left(L^{NS}_0 + \sum_n \left(n \eta_n \xi_n - (n+ \half) \b_{n+\half} \g_{n+\half} \right)\right) \Biggr\},
\eeq
That is the off-shell propagator for the Neveu-Schwarz sector
\beq \label{prons}
G_{NS} = T \int d \bar n \int D[X,P] D[\psi] D[\eta, \xi] D[\b,\g] e^{iS},
\eeq 

We observe that the path integral in Eq. (\ref{prons}), representing the off-shell propagator, is a typical path integral representation for the transition matrix element with Hamiltonian $H$:
\beq
H &=& L^{NS}_0 + L^{NS}_{gh}, \nn\\
L^{NS}_0 &=& \half \sum_n (P^2_n + n^2 X^2_n) + \sum_{n \ge 0} \left(n + \half\right) \left(\psi^\m_{-n-\half} \psi_{\m n+\half}\right), \nn\\.
L^{NS}_{gh} &=& \sum_n \left(n \eta_n \xi_n - 
\left(n+\half\right) \b_{n+\half} \g_{n+\half} \right).
\eeq 
Defining proper time $s=T\bar n$, we obtain the off-shell propagator in the NS sector as follows: 
\beq
G_{NS} &=& \int_0^\infty ds \langle X_f, \psi_f, \eta_f, \b_f \vert \exp\left\{,
-s(L^{NS}_0 + L^{NS}_{gh})\right\} \vert X_i, \psi_i, \eta_i, \b_i \rangle \nn\\
&=& \langle X_f, \psi_f, \eta_f, \b_f \vert \frac{1}{L^{NS}_0 + L^{NS}_{gh} - i\e},
\vert X_i, \psi_i, \eta_i, \b_i \rangle 
\eeq 

\subsection{Ramond Sector} 

We can construct the BRST generator $Q_R$ for the Ramond sector using the structural constants given in Eq. (\ref{constramond}). Similarly, 
\beq
Q_R &=& \eta_n L^R_n + \bar \eta_n \bar \Pi_n + \b_n F_n + \bar \b_n \bar \zeta_n - \frac{1}{2}(n-m) \eta_n \eta_m \xi_{n+m} \nn\\
&&+ \left(\frac{n}{2} -m \right) \eta_n \b_m \g_{n+m} - \b_n \b_m \xi_{n+m}
\eeq 
The BRST invariant action may be written as
\beq
S &=& \int^{\t_f}_{\t_i} d\t \Biggl\{\sum_n \left(P^\m_n \dot X_{\m n} 
+ \frac{i}{2} \psi^\m_{-n} \dot \psi_{\m n} + i \xi_n \dot \eta_n + i \g_n \dot \b_n\right) + \bar \Pi_0 \dot {\bar N}_0 + \bar \zeta_0 \dot {\bar \l}_0 \nn\\
&& + i \bar \xi_0 \dot {\bar \eta}_0 + i \bar \g_0 \dot {\bar \b}_0 - \sum_n (\bar N_n L^R_n + \bar \l_n F_n ) + \sum_{n,m} \Biggl((n-m) \bar N_n \eta_m \xi_{n+m} \nn\\
&& -\left(\frac{n}{2} -m \right) \bar N_n \b_m \g_{n+m} + \left(n-\frac{m}{2} \right) \bar \l_n \eta_m \g_{n+m} + 2 \bar \l_n \b_m \xi_{n+m} \Biggr) \nn\\
&& - \sum_n{}^\prime \left(\bar \Pi_n \bar N_n - \bar \zeta_n \bar \l_n - i \bar \eta_n \bar \xi_n - i \b_n \bar \g_n \right) \Biggr\},
\eeq 
We integrate $\bar \pi_n$ and $\bar \zeta_n$, and as a result, obtain the action parametrized by the modular and super-modular parameters in the covariant gauge
\beq
\int d \bar n d \n e^{iS}.
\eeq 
Integrations over the ghost variables $(\bar \eta_n, \bar \xi_n)$ $n\not=0$ and $(\bar \b_n, \bar \g_n)$, $n\not=0$ are trivial. We perform further integrations over the ghost zero modes $(\eta_0, \xi_0)$, $(\b_0,\g_0)$, $(\bar \eta_0, \bar \xi_0)$, and $(\bar \b_0, \bar \g_0)$ which do not induce a nontrivial factor in the measure, however, in contrast to the Neveu-Schwarz sector 

Furthermore, we reach the point where the off-shell propagator for the Ramond sector is explicitly evaluated.
\beq
S &=&\int^{\t_f}_{\t_i} d\t \Biggl\{\sum_n \left(P^\m_n \dot X_{\m n} 
+ \frac{i}{2} \psi^\m_{-n} \dot \psi_{\m n} \right) + i \sum_n{}^\prime (\xi_n \dot \eta_n + \g_n \dot \b_n ) \nn\\
&& - \bar n \left(L^R_0+ \sum_n{}^\prime n (\eta_n \xi_n - \b_n \g_n ) \right) -\n \left(F_0 - \sum_n{}^\prime \left(\frac{n}{2} \eta_n \g_n + 2 \b_n \xi_n \right)\right)\Biggr\},
\eeq 
after ghost zero modes were integrated. The precise definition of the path integral representation of the off-shell propagator in the Ramond sector is as follows:
\beq
G_R(X_f, \psi_f, \eta_f, \b_f; X_i, \psi_i, \eta_i, \b_i) = \int d \bar n d\n \int D[X,P] D[\psi] D[\eta,\xi] D[\b,\g] e^{i\bar S}.
\eeq 
Defining the proper time $s=T \bar n$ and its supersymmetric counterpart $\varrho = T \n$, we rewrite the path integral as 
\beq
G_R(X_f, \psi_f, \eta_f, \b_f; X_i, \psi_i, \eta_i, \b_i) &=& \int^\infty_0 ds \int d \varrho \langle X_f, \psi_f, \eta_f, \b_f \vert \exp\Bigl\{ -is(L^R_0 + L^R_{gh}) \nn\\ &&- i \varrho (F_0 + F_{gh}) \Bigr\} \vert X_i, \psi_i, \eta_i, \b_i \rangle,
\eeq 
We define 
\beq
L^R_0 &=& \frac{1}{2} \sum_n (P^2_n + n^2 X^2_n) + \half \sum_{n >0} n \left(\psi^m_{-n} \psi_{\m n} - \psi^\m_n \psi_{\m(-n)}\right), \nn\\
L^R_{gh} &=& \sum_n{}^\prime n \left(\eta_n \xi_n - \b_n \g_n \right) , \nn\\
F_0 &=& \sum_{n \ge 0} \frac{1}{\sqrt{2}} \left(P^\m_n + in X^\m_n \right) \psi_{\m n} + \sum_{n>0} \frac{1}{\sqrt{2}}\left(P^\m_n - in X^\m_n \right) \psi_{\m (-n)} , \nn, \\
F_{gh} &=& \sum_n{}^\prime \left(\frac{n}{2} \eta_n \g_n - 2 \b_n \xi_n \right).
\eeq 

Observing that 
\beq
\left[L^R_0+ L^R_{gh}, F_0 + F_{gh} \right] = 0 , 
\eeq 
and 
\beq
\left(F_0+F_{gh}\right)^2 = L^R_0 + L^R_{gh},
\eeq 
we can simply perform integration over the modular and super-modular parameters
\beq
G_R(X_f, \psi_f, \eta_f, \b_f; X_i, \psi_i, \eta_i, \b_i ) &=& \langle X_f, \psi_f, \eta_f, \b_f \vert \frac{F_0 + F_{gh}}{L^R_0 + L^R_{gh} - i\e} \vert X_i, \psi_i, \eta_i, \b_i\rangle \nn\\
&=& \langle X_f, \psi_f, \eta_f, \b_f \vert \frac{1}{F_0 + F_{gh}}\vert X_i, \psi_i, \eta_i, \b_i\rangle.
\eeq 

\section{Free Field Action for the NSR Open Super-String}

The expressions of the propagators obtained in the previous sections reveal the structure of the free-field actions for the NSR open super-string. The free-field action is constructed such that the off-shell propagators can be deduced in terms of the action through the quantum field theoretical expression. The free-string field action is given by: 
\beq
S = \int D[X,\psi,\eta,\b]_{NS} \Phi \left(L^{NS}_0 + L^{NS}_{gh} - i\e \right) \Phi +, 
\int D[X,\psi,\eta,\b]_{R} \Psi \left(F_0 + F_{gh} \right) \Psi.
\eeq 
We discuss the structure of action $S$ in detail. We may define the "creation" and "annihilation" operators
\beq
\psi^{\m\dag}_j = \psi^\m_{-j}, ~~r^\dag_j = \b_j, ~~ r_j = -\g_j, ~~ s^\dag_j = \g_{-j},~~ s_j=\b_{-j}
\eeq 
where $j$ can appropriately be a positive integer or half-integer. They satisfy the commutation relations of harmonic oscillators or anti-commutation relations of Grassman harmonic oscillators:
\beq
[\psi^\m_i, \psi^{\dag \n}_j ]_+= \eta^{\m\n}\d_{ij}, ~~ 
[r_i, r^\dag_j] = \d_{ij},~~ 
[s_i, s^\dag_j] = \d_{ij}
\eeq

The structure of field action becomes transparent in terms of creation and annihilation operators.
In the Neveu-Schwarz sector, the kinetic operator is given as
\beq
L^{NS}_0 &=& p^2 + N^{NS} - \frac{d}{16} \nn\\
&=& p^2 + \sum_{n >0} n a^{\dag \m}_n a_{\m n} + \sum_{n \ge 0} \left(n+ \half\right) \psi^{\dag \m}_{n+\half} \psi_{\m n+ \half} - \frac{d}{16}. \nn\\
L^{NS}_{gh} &=& N^{NS}_{gh} + \frac{1}{8} \nn\\
&=& \sum_{n>0} \sum_{i=1}^2 n a^{\dag i}_{\rm gh}{}_n a^{i}_{\rm gh}{}_n + \sum_{n \ge 0} \left(n+ \half\right) \left(r^\dag_{n+\half} r_{n+\half} + s^\dag_{n+\half} s_{n+\half} \right),
\eeq 
where $a^1_{\rm gh}{}_n, a^2_{\rm gh}{}_n$ may be written in terms of the usual $b$–$c$ ghost variables as 
\begin{subequations}
\beq
b_{zz}(\s) &=& \frac{b_0}{2} + \frac{1}{2} \sum_{n=1} \left(a_{1n}^{\text{gh}}e^{-in\s} -I a^{\dag\text{gh}}_{2n} e^{in\s}\right), \\
b_{\bar z\bar z}(\s) &=& \frac{b_0}{2} + \frac{1}{2} \sum_{n=1} \left(a_{1n}^{\text{gh}} e^{in\s} -i a^{\dag\text{gh}}_{2n} e^{-in\s} \right), \\
c^z(\s) &=& \frac{c_0}{2} + \frac{1}{2}\sum_{n=1} \left(a^{\dag\text{gh}}_{1n} e^{in\s} + ia_{2n}^{\text{gh}} e^{-in\s}\right), \\
c^{\bar z}(\s) &=& \frac{c_0}{2} + \frac{1}{2}\sum_{n=1} \left(a^{\dag\text{gh}}_{1n} e^{-in\s} 
+ ia_{2n}^{\text{gh}} e^{in\s}\right).
\eeq
\end{subequations}
They satisfy 
\beq
[a^{\dag i}_{\rm gh}{}_n, a^{j}_{\rm gh}{}_m]_+ = \d_{nm} \d^{ij}.
\eeq

In the Ramond sector, 
\beq
F_0 &=& p_\m \psi^\m_0 + \sum_{n >0} \sqrt{n} \left(a^{\dag\m}_n \psi_{\m n} + a^\m_n \psi^\dag_{\m n} \right), \nn\\
F_{\rm gh} &=& - \sum_{n >0} \left(\frac{n}{2}(a^{1\dag}_{\rm gh}{}_n + a^2_{\rm gh}{}_n s^\dag_n ) + 2(r^\dag_n a^{1}_{\rm gh}{}_n + s_n a^{2\dag}_{\rm gh}{}_n ) \right)
\eeq 
Because $\psi^\m_0$ satisfies the Clifford algebra
\beq
[\psi^\m_0, \psi^\n_0]_+ = \eta^{\m\n},
\eeq 
we may represent them by ten dimensional $G$ matrices
\beq
\psi^\m_0 = \G^\m.
\eeq 

We can define a mass operator $M$ to make the structure of the action in the Ramond sector more apparent. 
\beq
M &=& \sum_{n >0} \sqrt{n} \left(a^{\dag \m}_n\psi_{\m n} + a^\m_n \psi^\dag_{\m n} \right) ~~M_{gh} = F_{gh}, \nn\\
M^2 &=& \sum_{n >0} n \left( a^{\dag \m}_n A_{\m n} + \psi^{\dag \m}_n \psi_{\m n} \right) = N^R \nn\\
M^2_{gh} &=& \sum_{n>0} n \left( a^{\dag 1}_{\rm gh}{}_n a^{1}_{\rm gh}{}_n 
+ a^{\dag 2}_{\rm gh}{}_n a^{2}_{\rm gh}{}_n + r^\dag_n r_n + s^\dag_n s_n \right) = N^R_{gh}
\eeq 
Thus, the excited state of the super-string in the Neveu-Schwarz sector describes a particle with a mass given by the eigenvalues of mass operators 
$m^2 = N^{NS} + N^{NS}_{gh} -1/2$ at the critical dimensions $d=10$. The ground state of the super-string in the Neveu-Schwarz sector is tachyonic. This tachyonic ground state is removed by the GSO (Gliozzi-Scherk-Olive) projection operator \cite{GSO1977}. However, a Dirac-type particle that has a mass $m$, $m^2 = N^R + N^R_{gh}$, can realize an excited state of the super-string in the Ramond sector. The ground state in the Ramond sector has ten- dimensional spinor index with $E_{\rm ground} =0$. 

The GSO-projected free-field action is defined in terms of the total fermion number, $F$, which is BRST invariant.
\beq
F = \begin{cases} \sum_{n=1} \left(\psi^{\dag \m}_n \psi_{\m n} + r^\dag_n r_n + s^\dag_n s_n \right), ~~~~~~& \text{Ramond sector} \nn\\,
\sum_{n=0}\left(\psi^{\dag \m}_{n+\half} \psi_{\m n+\half} + r^\dag_{n+\half} r_{n+\half} + s^\dag_{n+\half} s_{n+\half} \right) , ~~~~~~& \text{Neveu-Schwarz sector}
\end{cases}
\eeq 
by
\beq
S &=& \half \int [X,\psi,\eta,\b]_{NS} \Phi \left(1 -(-1)^F\right)\left(- \p^\m \p_\m + N^{NS} + N^{NS}_{gh} - \frac{1}{2} \right) \Phi \nn, \\
&&+\half \int [X,\psi,\eta,\b]_{R} \Psi \left(1+ \G^{11}(-1)^F\right) \left(\p_\m \G^\m + M + M_{gh} \right) \Psi 
\eeq

\section{Conclusions and Discussions}

We studied the Polyakov string path integral of an NSR open superstring on a strip. The evaluation of the path integral was performed in a manifestly covariant manner.
This was achieved by choosing the covariant gauge conditions that were imposed on the metric and its super-symmetric counterpart (gravitino) on the strip. 
The NSR open super-string had two sectors, depending on the periodicity of the fermion field: the NS sector with a periodic condition and the Ramond sector with an anti-periodic sector. In the proper time gauge, the NS sector had modular
(proper-time) parameter, and the R sector had both modular and supermodular parameters. Integration of modular and super-modular parameters yielded string propagators in both sectors.
We demonstrated that the path integrals represent field-theoretical off-shell string propagators, which may be obtained from string free-field actions \cite{Terao96,Ohta86,deAlwis86,Date86}. Hence, this study revealed the connection between geometric and algebraic approaches. 

This study can be extended in various directions. The immediate extension may be to evaluate the Polyakov string-path integrals of the NSR super string over the Riemann surface, describing three and four interacting strings. Accordingly, we will obtain full string vertex operators, just as in the case of bosonic string theory. The (super)string theory in the proper-time gauge is deformable to Witten's (super)string. However, in the proper-time gauge, it is easy to deal with string interactions of an arbitrary number of strings because the Riemann surfaces in the 
proper time gauges are free of conical singularity, which is the main obstacle in evaluating the scattering amplitudes of higher string interactions. In the presence of conical singularity, it is difficult to prove the local (non-Abelian) gauge invariance of scattering amplitudes on multiple $Dp$-branes. It is also unclear whether the extended BRST impedance is violated by higher-order string interactions. As in the case of bosonic string theory, we expect that the scattering amplitudes of the NSR super-string with three and four strings can be calculated in the proper-time gauge without difficulty.

It is important to derive conventional QFT from corresponding super-string accurately. Otherwrise, it may be difficult to systematically calculate the stringy corrections. However, the fermionic action is introduced by hand to compatible with the 
spectrum of the free superstring. In the case of the interacting theory, things get even worse if we employ the (super)-conformal field theory. Unlike the theory in the proper-time gauge, the conventional NSR superstring, based on the 
conformal field theory possesses zero modes. Because the conformal field theory for the NSR superstring is defined on a sphere in contrast to the theory in the proper-time gauge, which is defined on a cylindrical surface. To deal with these zero modes, we need to insert some  (picture changing) operators at the midpoint of the string. Yet this midpoint insertion is difficult to be defined properly. In contrast to the conventional approach, in the proper-time gauge approach the interacting NSR superstring, which is free of this problem, may be derived from the Polyakov string path integral	consistently.
This work is important as a preliminary work along this direction. 

\vskip 1cm

\begin{acknowledgments}

TL was supported by the National Research Foundation of Korea(NRF) grant, funded by the Korean government (MSIT) (2021R1F1A106299311).
Part of this study was conducted during the author's visit to the APCTP (Korea, Pohang). This study has been worked also with the support of a research grant of Kangwon National Universtiy in 2019. 
\end{acknowledgments}




\end{document}